\documentclass[final,5p,sort&compress]{elsarticle}
\usepackage[applemac]{inputenc}
\usepackage{graphicx}
\usepackage{xcolor}
\usepackage{amsmath}
\newcommand{\bea}{\begin{eqnarray}}
\newcommand{\eea}{\end{eqnarray}}
\journal{Physics Letters B}
\begin{document}
\begin{frontmatter}
\title{Freeze-out conditions from net-proton and net-charge fluctuations at RHIC}
\author{Paolo Alba$^a$, Wanda Alberico$^a$, Rene Bellwied$^b$}
\author{Marcus Bluhm$^{a,c}$}
\author{Valentina Mantovani Sarti$^{a}$, Marlene Nahrgang$^{d,e}$, Claudia Ratti$^a$}
\address{$^a$ Department of Physics, Torino University and INFN, Sezione di Torino, via P. Giuria 1, 10125 Torino, Italy\\
$^b$ Department of Physics, University of Houston, Houston, TX 77204, USA\\
$^c$ Department of Physics, North Carolina State University, Raleigh, NC 27695, USA\\
$^d$ Department of Physics, Duke University, Durham, NC 27708-0305, USA\\
$^e$ Frankfurt Institute for Advanced Studies (FIAS), Ruth-Moufang-Str. 1, 60438 Frankfurt am Main, Germany}
\begin{abstract}
We calculate ratios of higher-order susceptibilities quantifying fluctuations in the number of net protons and in the net-electric charge using the Hadron Resonance Gas (HRG) model. We take into account the effect of resonance decays, the kinematic acceptance cuts in rapidity, pseudo-rapidity and transverse momentum used in the experimental analysis, as well as a randomization of the isospin of nucleons in the hadronic phase. By comparing these results to the latest experimental data from the STAR collaboration, we determine the freeze-out conditions from net-electric charge and net-proton distributions and discuss their consistency.
\end{abstract}
\end{frontmatter}

\section*{Introduction}

Significant theoretical activity, aimed at understanding the properties of matter under extreme conditions,
has been triggered recently by the heavy-ion collision experiments conducted at RHIC and the LHC, in which the
deconfined phase of QCD matter, the Quark-Gluon Plasma, is created. The transition from the
hadronic to the deconfined, partonic phase is an analytic crossover at zero baryo-chemical potential $\mu_B$~\cite{Aoki:2006we} with a transition temperature $T_c$ determined in lattice QCD simulations~\cite{Tctrilogy}. This crossover feature also extends to small values of $\mu_B$. The possibility that the transition becomes first-order for larger $\mu_B$, which would imply the existence of a critical point, is currently investigated by the beam energy scan program at RHIC, soon to be followed by the CBM experiment at FAIR.

Event-by-event fluctuations of the net-electric charge and net-baryon number, which are conserved charges of QCD, are expected to become large near a critical point~\cite{Berges:1998rc,Halasz:1998qr}: for this reason, they have been proposed as ideal observables to verify its existence and to determine its position in the QCD phase diagram~\cite{Stephanov:1998dy,Stephanov:1999zu,Gavai:2008zr,Endrodi:2011gv}. Experimental results for these measures were recently reported for several collision energies~\cite{Adamczyk:2013dal,charge,McDonald:2012ts,Sahoo:2012bs}.
In addition, as a consequence of the increasing precision achieved in the numerical simulations, it is becoming possible to extract the chemical freeze-out parameters (i.e. freeze-out temperature $T_{ch}$ and corresponding baryo-chemical potential $\mu_{B,ch}$) from first principles, by comparing the measured fluctuation observables to corresponding susceptibility ratios calculated in lattice QCD~\cite{Karsch:2012wm,Bazavov:2012vg,Borsanyi:2013hza,Mukherjee:2013lsa,xxx}. When compared to experimental data from heavy-ion collisions, present lattice simulations have, however, their limitations: they cannot take the experimental acceptance cuts into account and they are available only for small chemical potentials. As a consequence of the latter, only the lowest order susceptibilities are available on the lattice at finite $\mu_B$. Moreover, the experimental restriction of net-baryon to net-proton number measurements cannot be realized on the lattice.

Recently, fluctuation observables have been investigated in transport approaches~\cite{Schuster:2009jv,Sahoo:2012wn} as well as in various baseline studies within the HRG model~\cite{Begun:2006jf,Karsch:2010ck,Fu,Garg:2013ata,Nahrgang:2014fza}. Calculations based on the HRG model in chemical equilibrium reproduce the equilibrium lattice QCD results for the susceptibilities and their ratios in the hadronic phase reasonably well~\cite{Borsanyi:2013hza,Borsanyi:2011sw}. Furthermore, the model allows to expand the range of $\mu_B$-values and consequently to calculate ratios of higher order susceptibilities at finite $\mu_B$, as well as to implement kinematic acceptance cuts in rapidity $y$, pseudo-rapidity $\eta$ and transverse momentum $p_T$, thus, providing a valuable tool to extract the freeze-out conditions from the experimental data.
In the past, statistical hadronization models (SHMs) have been used to analyze experimental data on particle production by comparing the data to thermal abundances calculated in HRG model approaches for all collision energies ranging from AGS to the LHC, see e.g.~\cite{BraunMunzinger:2003zd,Becattini:2005xt,Manninen:2008mg,Cleymans:2005xv,Andronic:2011yq} and references therein. The freeze-out temperatures determined in this way lie, however, at the upper limit of the uncertainty band of the lattice QCD results for $T_c$~\cite{Tctrilogy}, which is most pronounced for the highest collision energies.

In this paper, we calculate ratios of susceptibilities quantifying fluctuations in the number of net protons and in the net-electric charge within the HRG model: our study includes the effects of resonance decays and isospin randomization for the nucleons, as well as kinematic acceptance cuts in agreement with the experimental analysis. We extract new freeze-out points in the ($T,\mu_B$)-plane of QCD matter and compare them with the freeze-out curve determined in Ref.~\cite{Cleymans:2005xv}. Finally, we discuss the issue of the consistency between the freeze-out parameters obtained from the analysis of net-electric charge and net-proton fluctuations. We find that it is possible to use a combined analysis of the lowest-order cumulants of the net-electric charge and the net-proton distributions in order to extract common freeze-out conditions $T_{ch}$ and $\mu_{B,ch}$. Possible limitations of this method are addressed.

\section*{The HRG in partial chemical equilibrium}

The HRG model provides a suitable description of the bulk properties of hadronic matter in thermal and
chemical equilibrium, see e.g.~\cite{Karsch03,Tawfik05}. In the thermodynamic limit, the pressure as a function of temperature $T$ and all hadron chemical potentials $\mu_k$ is given by
\begin{multline}
p(T,\{\mu_k\}) =  \sum_k (-1)^{B_k+1} \frac{d_kT}{(2\pi)^3} \int {\rm d}^3\vec{p} \,\,\ln
\Big[1+\\		(-1)^{B_k+1} \exp({-(\sqrt{\vec{p}^{\,2}+m_k^2}-\mu_k)/T})\Big] \,,
\label{equ:pressureHRG}
\end{multline}
where the sum is taken over all hadronic states $k$, including resonances, in the model (baryons and anti-baryons being summed independently). In Eq.~(\ref{equ:pressureHRG}), $d_k$ and $m_k$ denote the degeneracy factor and the mass, respectively, and $\mu_k=B_k\mu_B+Q_k\mu_Q+S_k\mu_S$ is the chemical potential of the species $k$ in chemical equilibrium. $B_k$, $Q_k$ and $S_k$ are the respective quantum numbers of baryon charge, electric charge and strangeness, while $\mu_B$, $\mu_Q$ and $\mu_S$ denote the chemical potentials associated with the net densities of baryon number, $n_B$, electric charge, $n_Q$, and strangeness, $n_S$, respectively.
The particle number density $n_k=N_k/V$ of species $k$ follows from the thermodynamic identity
$n_k\!=\!(\partial p/\partial\mu_k)_T$ as
\begin{multline}
\label{equ:particledensity}
n_k(T,\mu_k)=\frac{d_k}{(2\pi)^3}\int {\rm d}^3\vec{p}\,\times\\		
\times \frac{1}{(-1)^{B_k+1}+\exp{((\sqrt{\vec{p}^{\,2}+m_k^2}-\mu_k)/T})}
\end{multline}
so that the above net densities are given by $n_X=\sum_k X_k n_k$ for $X=B,Q,S$.

The actual conditions present in a heavy-ion collision are, however, more complex: first, the chemical potentials
$\mu_B$, $\mu_Q$ and $\mu_S$ are not independent, but related to $T$ as well as to
each other via the conditions
\begin{align}
\label{conditions}
n_{S}(T,\mu_B,\mu_Q,\mu_S)&=0
\nonumber\\
n_{Q}(T,\mu_B,\mu_Q,\mu_S)&=0.4\,n_B(T,\mu_B,\mu_Q,\mu_S) \,.
\end{align}
Here, the factor $0.4$ in Eq.~(\ref{conditions}) accounts approximately for the ratio of protons to baryons
in the colliding nuclei, while $n_S=0$ reflects the initial net-strangeness content.

Second, during its expansion the created matter does not maintain chemical equilibrium, since the time scales for the necessary inelastic scatterings among the hadrons are typically much longer than the duration of the hadronic stage~\cite{Teaney02}.
While hadrons are assumed to be formed in chemical equilibrium at the transition temperature $T_c$, for not too small temperatures $T\leq T_{ch}\leq T_c$ only the particle-number preserving interactions mediated by hadronic resonance decay and regeneration (e.g. $\pi\pi\to\rho\to\pi\pi$, $K\pi\to K^*\to K\pi$, $p\pi\to\Delta\to p\pi$ etc) continue to occur with sufficient rate.
The hadronic matter is, therefore, in a state of partial chemical equilibrium~\cite{Bebie92}, because the resonances are still in chemical equilibrium with their decay products, whereas the multiplicity ratios of hadrons, which are stable against strong decay during the hadronic stage, are frozen out at $T_{ch}$. The chemical potentials of the resonances $R$ become functions of the chemical potentials of all stable hadrons $h$ via $\mu_R=\sum_h\mu_h\langle n_h\rangle_R$ and, consequently, the final particle number of a hadron species $h$ is given by $\hat{N}_h=N_h+\sum_R N_R\, \langle n_h\rangle_R$, where the sum runs over all $R$ decaying into $h$, $N_h$ and $N_R$ denote the primordial particle numbers of $h$ and $R$, and $\langle n_h\rangle_R$ gives the average number of $h$ produced in the decay of $R$. As discussed in Ref.~\cite{Nahrgang:2014fza}, this connection can easily be extended toward higher-order susceptibilities in order to study the average influence of resonance decays on the fluctuations in the final particle numbers.

In this paper we consider, in line with Ref.~\cite{Bluhm:2013yga}, a HRG model containing states up to a mass of $2$~GeV as,
for example, listed in the Particle Data Book~\cite{PDG12}. We consider as stable particles the
mesons $\pi^0$, $\pi^+$, $\pi^-$, $K^+$, $K^-$, $K^0$, $\overline{K}^0$ and $\eta$ and the baryons $p$, $n$,
$\Lambda^0$, $\Sigma^+$, $\Sigma^-$, $\Xi^0$, $\Xi^-$ and $\Omega^-$ as well as their corresponding
anti-baryons. This implies that, as in the experimental analysis, feed-down from weak decays is explicitly excluded in our approach.

When applying the experimental acceptance cuts, we modify the HRG model integrals in the following way:
\begin{multline}
\label{equ:particledensitycuteta}
n_k(T,\mu_k)=\frac{d_k}{4\pi^2}\int_{-\eta_{\rm MAX}}^{\eta_{\rm MAX}} {\rm d}\eta\int_{p_T^{\rm MIN}}^{p_T^{\rm MAX}} {\rm d}p_T\times
 \\
 \times \frac{p_T^2{\rm Cosh}[\eta]}{(-1)^{B_k+1}+\exp({(\sqrt{p_T^{\,2}{\rm Cosh}[\eta]^2+m_k^2}-\mu_k)/T})	}
\end{multline}
in the case of cuts on the pseudo-rapidity $\eta$ (for net-electric charge, i.e. for all charged particles) and
\begin{multline}
\label{equ:particledensitycuty}
n_k(T,\mu_k)=\frac{d_k}{4\pi^2}\int_{-y_{\rm MAX}}^{y_{\rm MAX}} {\rm d}y\int_{p_T^{\rm MIN}}^{p_T^{\rm MAX}} {\rm d}p_T\times
 \\
 \times \frac{p_T\sqrt{p_T^2+m_k^2}{\rm Cosh}[y]}{(-1)^{B_k+1}+\exp({({\rm Cosh}[y]\sqrt{p_T^{\,2}+m_k^2}-\mu_k)/T})}
\end{multline}
in the case of cuts on the rapidity $y$ (for net protons). Due to the setup of the HRG model, we can only apply cuts on the momenta of the primordial resonances and of the primordial stable hadrons, but not on those of the products of resonance decays. It has, however, been estimated that the effect of acceptance cuts in rapidity on the decay daughters should be in the percent range~\cite{Jeon:1999gr}.

Furthermore the effect of radial flow on the measured particle species in a limited transverse momentum range, which might have been particularly important for the net charge since there a composite of particle species with different mass is considered, has been estimated taking into account the blastwave parameters extracted by STAR, and found to be negligible due to the rather wide $p_T$-window of the measurement (0.2-2 GeV/c). This result is in agreement with a similar previous study~\cite{Garg:2013ata}.

\section*{Results}

In the following, we compare our HRG model calculations with the efficiency corrected experimental results for the most central collisions ($0-5\%$) published by the STAR collaboration for net-proton\footnote{The efficiency-corrected data for the lowest cumulant ratio ($\chi_2/\chi_1$) for net-protons can be found on the public STAR webpage.}~\cite{Adamczyk:2013dal} and net-electric charge fluctuations~\cite{charge}.
The susceptibilities of conserved charges are defined as
\begin{equation}
\chi_{lmn}^{BSQ}=\frac{\partial^{\,l+m+n}(p/T^4)}{\partial(\mu_{B}/T)^{l}\partial(\mu_{S}/T)^{m}\partial(\mu_{Q}/T)^{n}} \,.
\end{equation}
Their relationship with the central moments of the conserved charge multiplicity distributions is
\begin{align}
\mathrm{ mean:}&~~M=\langle N\rangle= VT^3\chi_1\nonumber\,,\\
\mathrm{ variance:}&~~\sigma^2=
\langle (\delta N)^2\rangle=VT^3\chi_2
\nonumber\,,
\\
\mathrm{ skewness:}&~~S=\frac{\langle (\delta N)^3\rangle}{\sigma^3}=\frac{VT^3\chi_3}{(VT^3\chi_{2})^{3/2}}
\nonumber\,,
\\
\mathrm{kurtosis:}&~~\kappa=\frac{\langle (\delta N)^4\rangle}{\sigma^4}-3=\frac{VT^3\chi_4}{(VT^3\chi_{2})^{2}}\,\,,
\label{cenmom}
\end{align}
where $\delta N=N-\langle N\rangle$. From the quantities in Eq.~(\ref{cenmom}) the following, to leading order volume-independent, ratios can be defined:
\begin{align}
\sigma^2/M&=\chi_2/\chi_1\,,~~~~~~~~~~~~S\sigma=\chi_3/\chi_{2}\,,
\nonumber\\
\kappa\sigma^2&=\chi_4/\chi_{2}\,,~~~~~~S\sigma^3/M=\chi_3/\chi_1\,.
\nonumber
\label{moments}
\end{align}

We calculate the net-proton fluctuations according to the method presented in Ref.~\cite{Nahrgang:2014fza}, where besides kinematic acceptance cuts also resonance decays and regeneration below the chemical freeze-out are taken into account. While resonance decays feed the distributions of the primordial protons and anti-protons, the regeneration of resonances leads to a randomization of the nucleon isospin: the dominant process is the regeneration of $\Delta(1232)$-resonances from the scatterings of nucleons with thermal pions. Subsequently, these $\Delta$-resonances decay into either the same or the opposite isospin state, where neutrons are, however, not detected experimentally. Consequently, additional fluctuations in the net-proton number arise, which we include based on the formalism by Kitazawa and Asakawa (KA)~\cite{Kitazawa:2011wh,Kitazawa:2012at}.

Net-electric charge fluctuations are calculated based on the most abundant charged particles, namely pions, kaons, and protons as well as their anti-particles. Also here, primordial distributions are fed by resonance decays, but corrections similar to the KA-corrections for the net-proton number are not needed, because processes via intermediate resonances conserve electric charge.

While the application of the experimental acceptance cuts is straightforward in the net-proton case (where $0.4$~GeV/c~$<p_T<0.8$~GeV/c and $|y|<0.5$), it is more difficult in the case of net-electric charge. Here, the general cuts are $0.2$~GeV/c~$<p_T<2$~GeV/c and $|\eta|<0.5$, but in order to suppress spallation protons, all protons (and anti-protons) with $p_T<0.4$~GeV/c are removed in the experimental analysis. Due to correlated resonance decay contributions to (anti-)protons and pions or kaons, e.g. $\Delta^{++}\to p+\pi^+$ or $\Lambda^0(1520)\to p+K^-$, which are given by a single integral in the HRG model calculation, we cannot cut the resonance contribution to the (anti-)protons in the same way without also affecting the contributions to the pions and kaons. We thus apply the lower $p_T$-cut of $0.4$~GeV/c only to the primordial protons and anti-protons.

In order to extract the freeze-out temperature and baryo-chemical potential for each collision energy, we have simultaneously analyzed two
experimentally measured susceptibility ratios. With the resulting freeze-out conditions ($T_{\rm ch}$, $\mu_{B, {\rm
ch}}$) we can calculate the remaining susceptibility ratios, which gives us a cross-check on the reliability of the determined freeze-out
parameters. The large experimental uncertainties in the higher-order susceptibility ratios of the net-electric charge $\chi_3/\chi_2$ and $\chi_4/\chi_2$ do not allow to meaningfully constrain the freeze-out temperature and baryo-chemical potential from net-electric charge fluctuations alone. Moreover, for the net protons, as already noted in Ref.~\cite{Nahrgang:2014fza}, it is not possible to simultaneously
reproduce $\sigma^2/M$ and $S\sigma$ for all beam energies: this might point at a limitation in our approach, for example, due to an overestimate of the KA-corrections, which maximize the impact of isospin randomization. Several other effects that might impact the higher order moments have also not yet been considered, such as volume fluctuations \cite{Skokov:2012ds}, exact (local) charge conservation \cite{Schuster:2009jv,Bzdak:2012an} or repulsive van-der-Walls forces among hadrons \cite{Fu}. Finally, the discrepancy in particular at larger $\mu_B$, could also hint at the onset of chiral critical fluctuations in the higher moments (skewness and above)~\cite{Redlich:2012xf,Skokov:2011rq}.

We therefore perform, first, a combined analysis of the ratios with the smallest experimental uncertainty, namely $\sigma^2/M$ for net-electric charge and for net protons. In addition, we consider an alternative analysis using higher-order cumulants, namely $\sigma^2/M$ for net-electric charge and $S\sigma$ for net protons, and discuss the difference in the extracted freeze-out parameters between these two choices.

\begin{figure}[h!]
\centering
\includegraphics[width=0.5\textwidth]{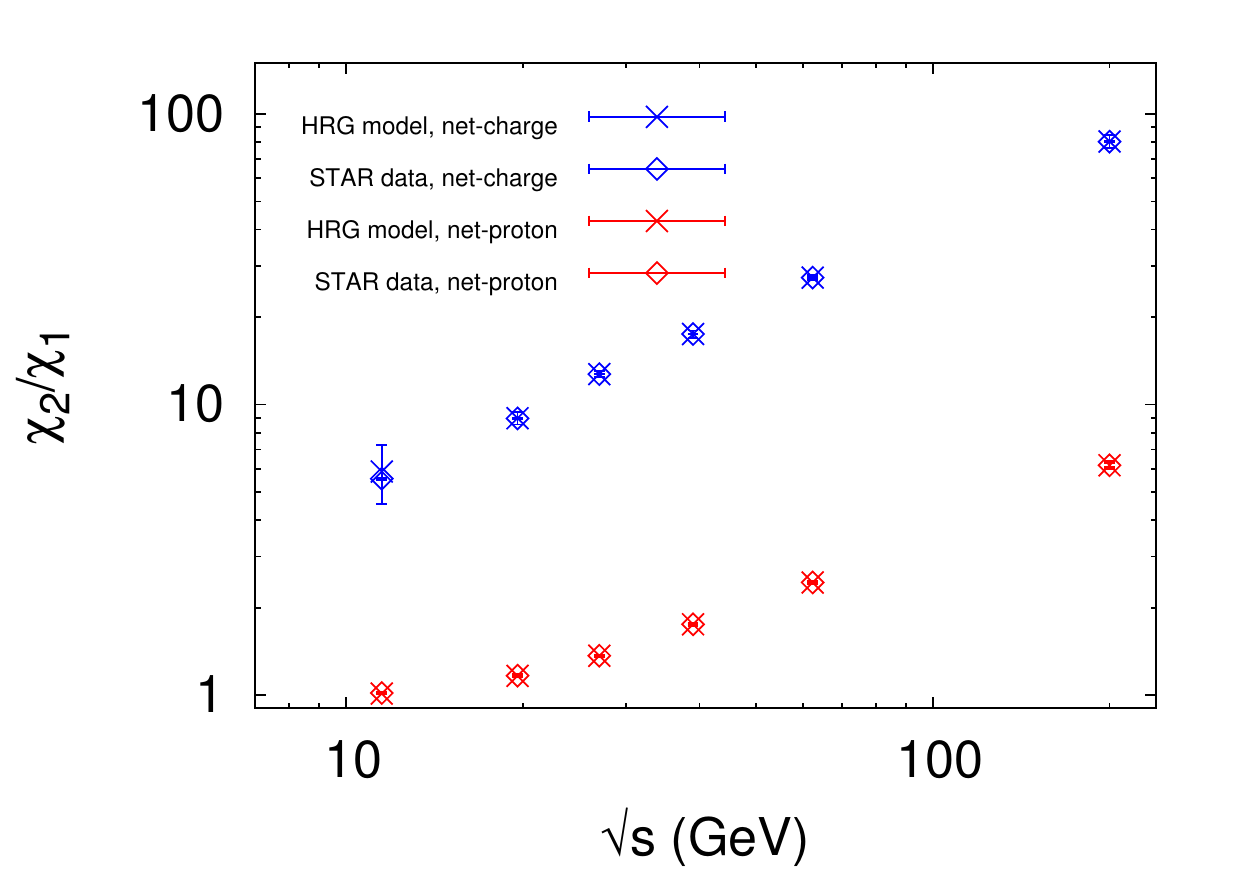}
\caption[]{\label{fig1} (Color online) Comparison between HRG model results and experimental data for the most central collisions ($0-5\%$) (from Refs.~\cite{Adamczyk:2013dal,charge}) for $\sigma^2/M$ of net-electric charge (blue, upper symbols) and net protons (red, lower symbols). The experimental data have been used in the HRG model in order to extract a freeze-out temperature and baryo-chemical potential for each collision energy.}
\end{figure}

In Fig.~\ref{fig1}, we show the experimental data as a function of collision energy per nucleon pair $\sqrt{s}$ (from Refs.~\cite{Adamczyk:2013dal,charge}) together with our results for the first choice of fluctuation observables, i.e. the combined $\sigma^2/M$ datasets. We find that it is possible to extract, for each collision energy, a freeze-out temperature and baryo-chemical potential, which allow to simultaneously reproduce the ratios of the lowest-order susceptibilities for
net protons and net-electric charge. The smallest collision energy we consider is $\sqrt{s}=11.5$ GeV: below
this energy we expect that the isospin randomization is not realized~\cite{Nahrgang:2014fza,Kitazawa:2011wh,Kitazawa:2012at}.
We note that for the determination of these freeze-out parameters the inclusion of the KA-corrections for $\sigma^2/M$ of net protons, in accordance with Ref.~\cite{Nahrgang:2014fza}, is essential.

\begin{figure}[h!]
\centering
\begin{minipage}{0.5\textwidth}
\includegraphics[width=0.9\textwidth]{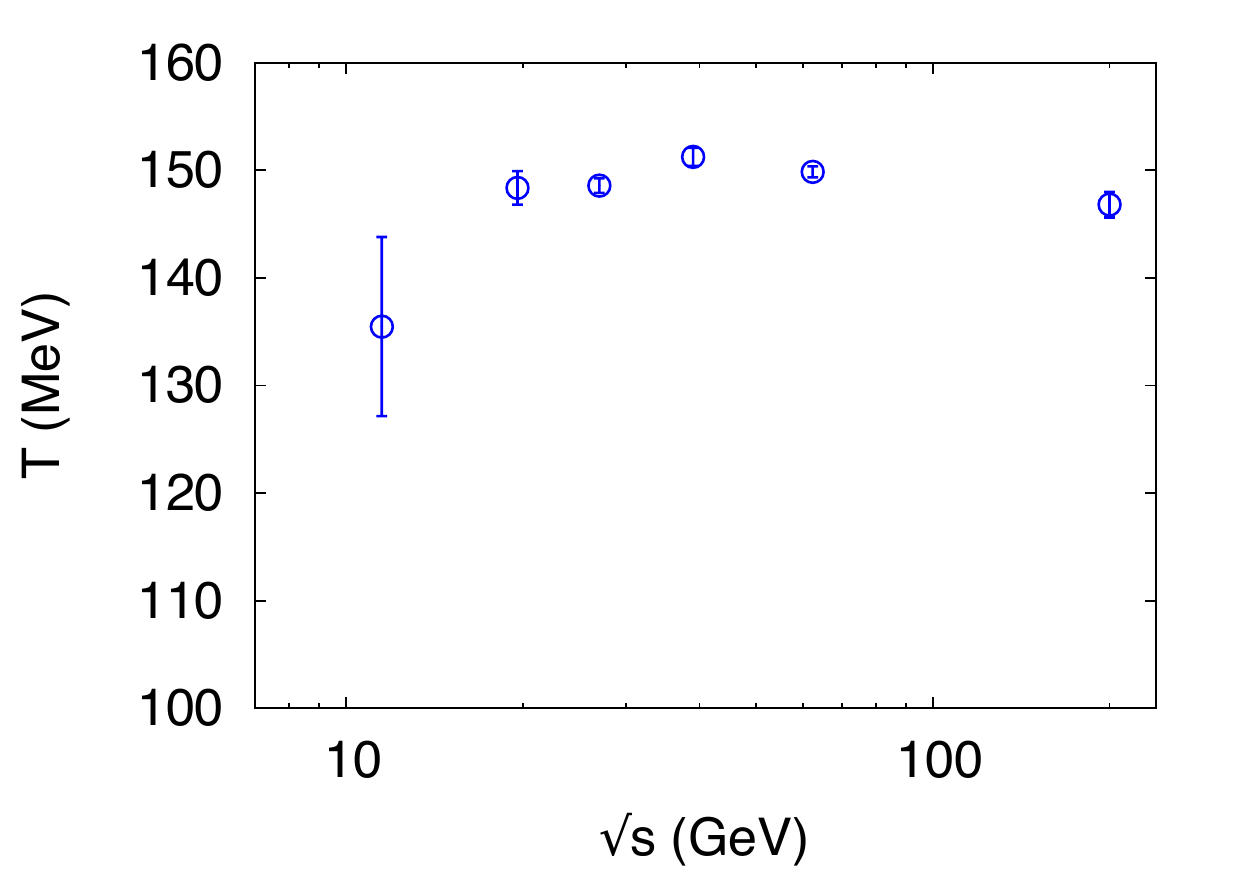}
\vspace{-.2cm}
\includegraphics[width=0.9\textwidth]{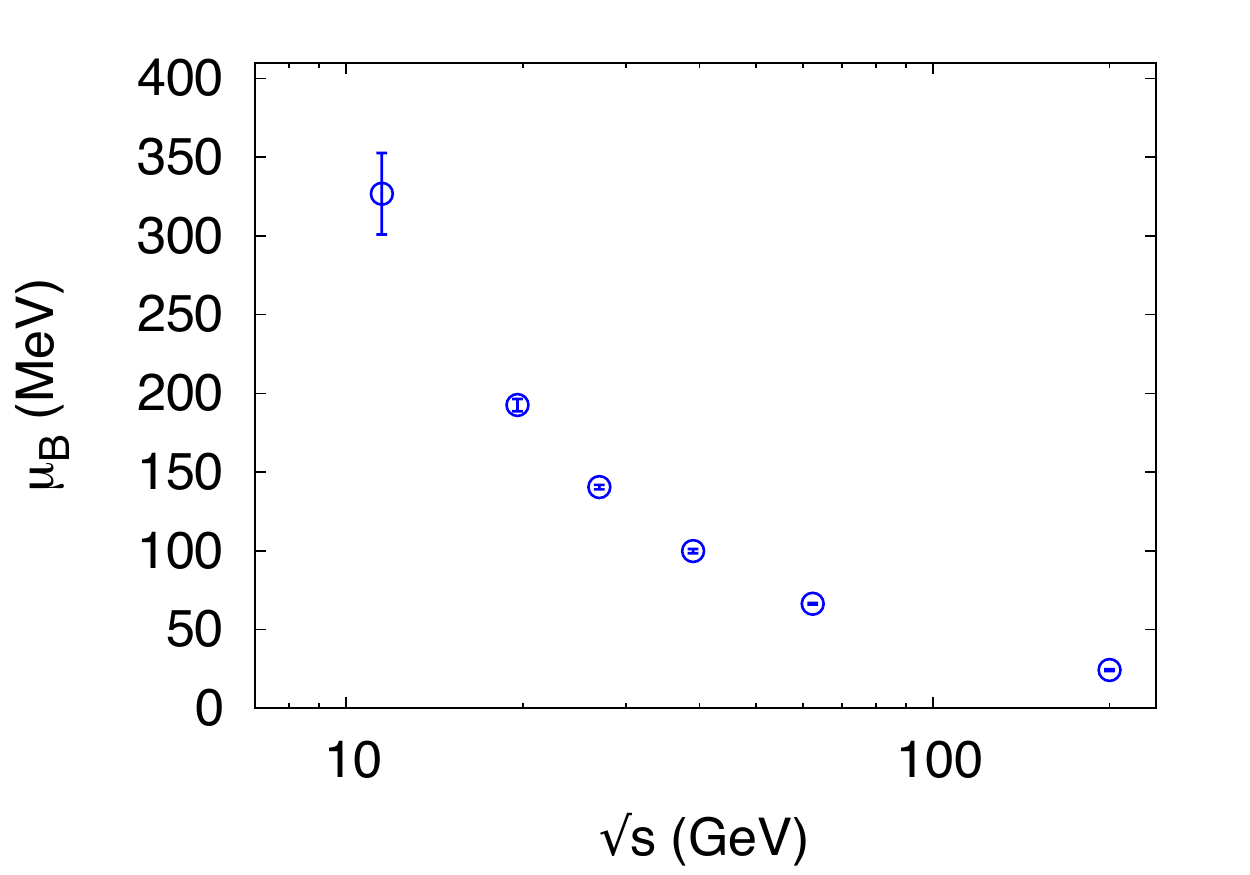}
\end{minipage}
\caption[]{\label{fig2} Freeze-out temperature (upper panel) and baryo-chemical potential (lower panel) as functions of the collision energy, extracted from the data in Fig.~\ref{fig1}. The corresponding values are listed in Table~\ref{tab1}.}
\end{figure}

\begin{figure}[h!]
\centering
\includegraphics[width=0.5\textwidth]{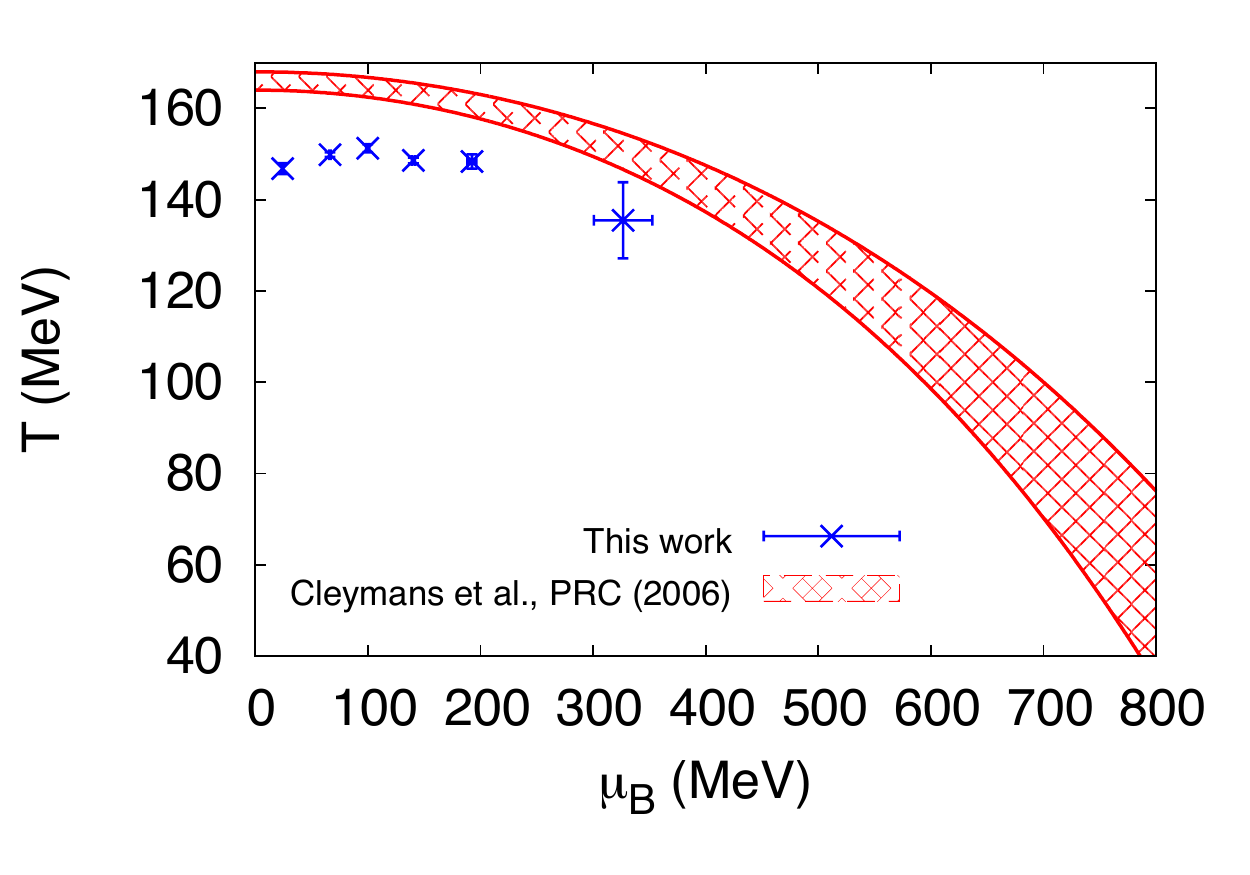}
\caption[]{\label{fig3} (Color online) Freeze-out parameters in the ($T-\mu_B$) plane: comparison between the curve obtained in Ref.~\cite{Cleymans:2005xv} (red band) and the values obtained from the combined analysis of $\sigma^2/M$ for net-electric charge and net protons (blue symbols) presented here.}
\end{figure}

In Fig.~\ref{fig2}, we show the freeze-out temperature (upper panel) and baryo-chemical potential (lower panel) corresponding to this set of analyzed cumulant ratios, as functions of $\sqrt{s}$. The precision in the experimental results allows a rather precise determination of these parameters. The error bars shown in Fig.~\ref{fig2} are based on HRG model calculations using the upper and lower uncertainty limits in the experimental data. Our values for $T_{ch}$ are lower than those found in Ref.~\cite{Cleymans:2005xv}: even for the highest RHIC energies, our results are close to the lower bound for $T_c$ determined in lattice QCD simulations~\cite{Tctrilogy}.
This is evident in Fig.~\ref{fig3}, where we show a comparison between the freeze-out points in the ($T-\mu_B$) plane obtained in the present analysis and the curve of Ref.~\cite{Cleymans:2005xv}.

\begin{figure}[h!]
\centering
\begin{minipage}{0.5\textwidth}
\includegraphics[width=0.5\textwidth]{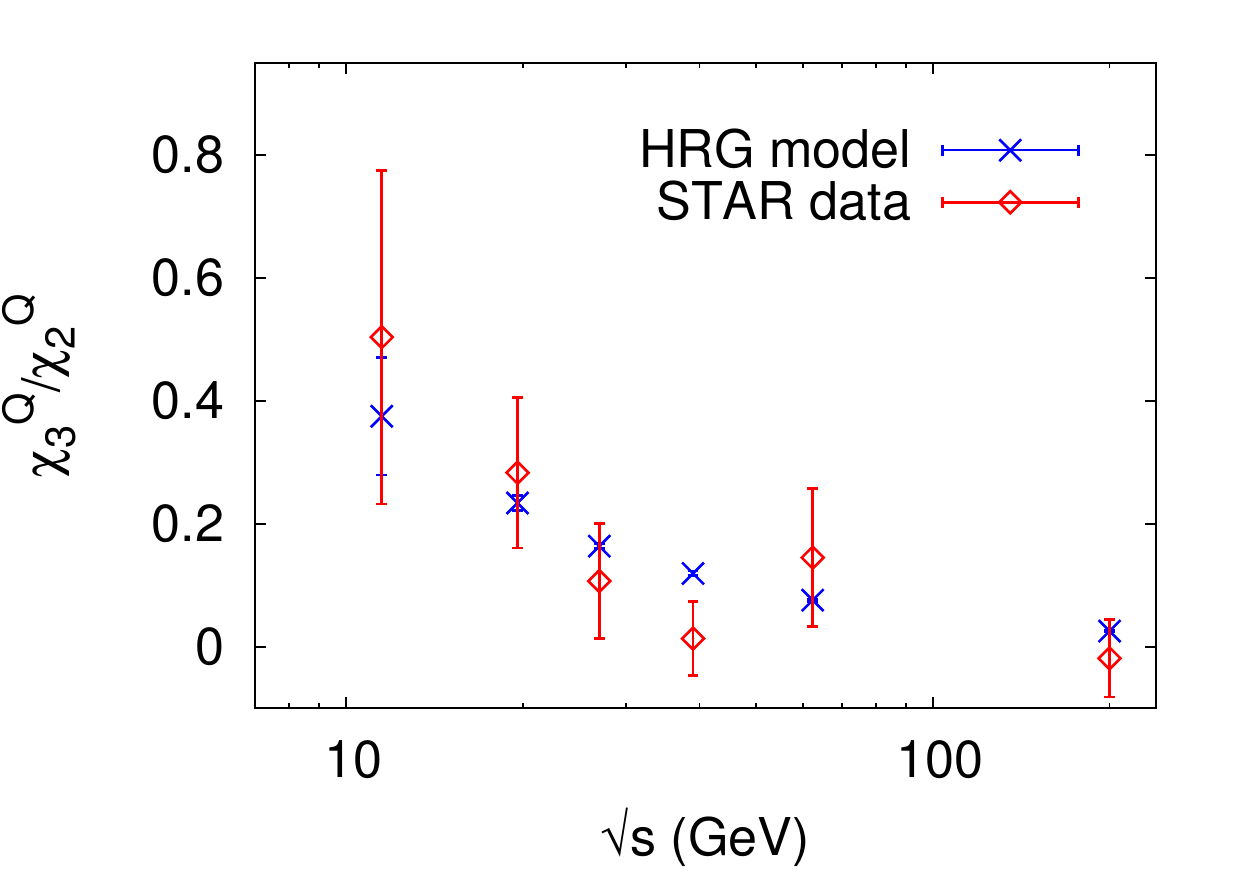}
\includegraphics[width=0.5\textwidth]{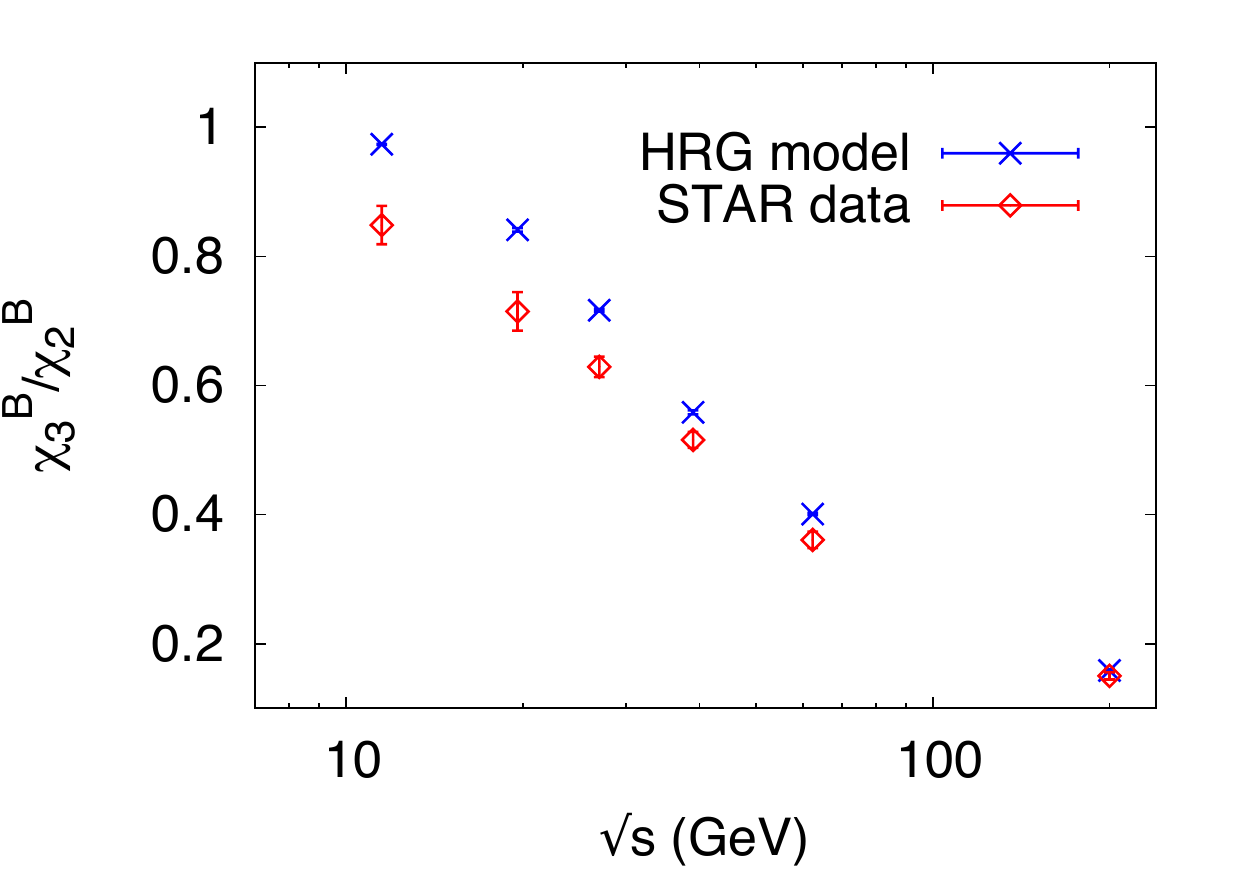}
\includegraphics[width=0.5\textwidth]{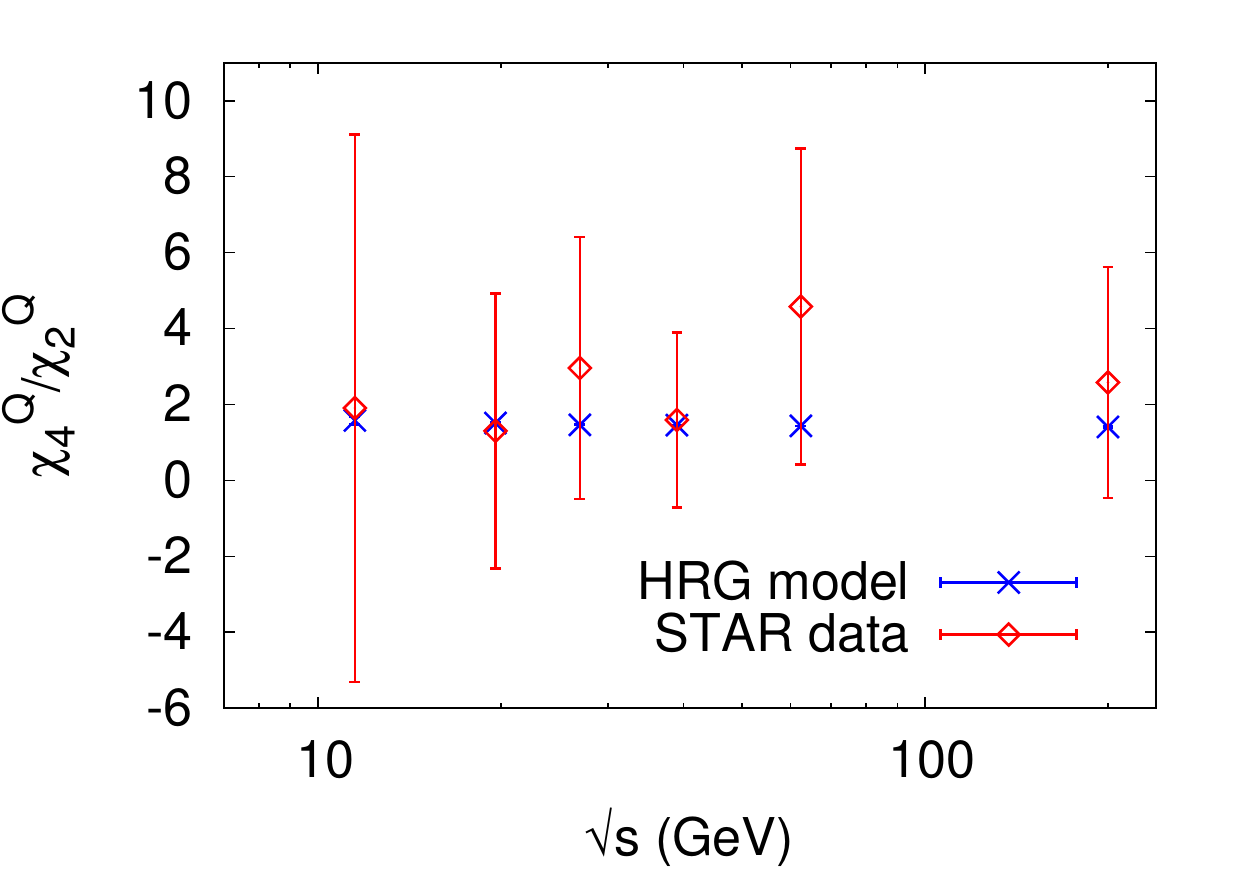}
\includegraphics[width=0.5\textwidth]{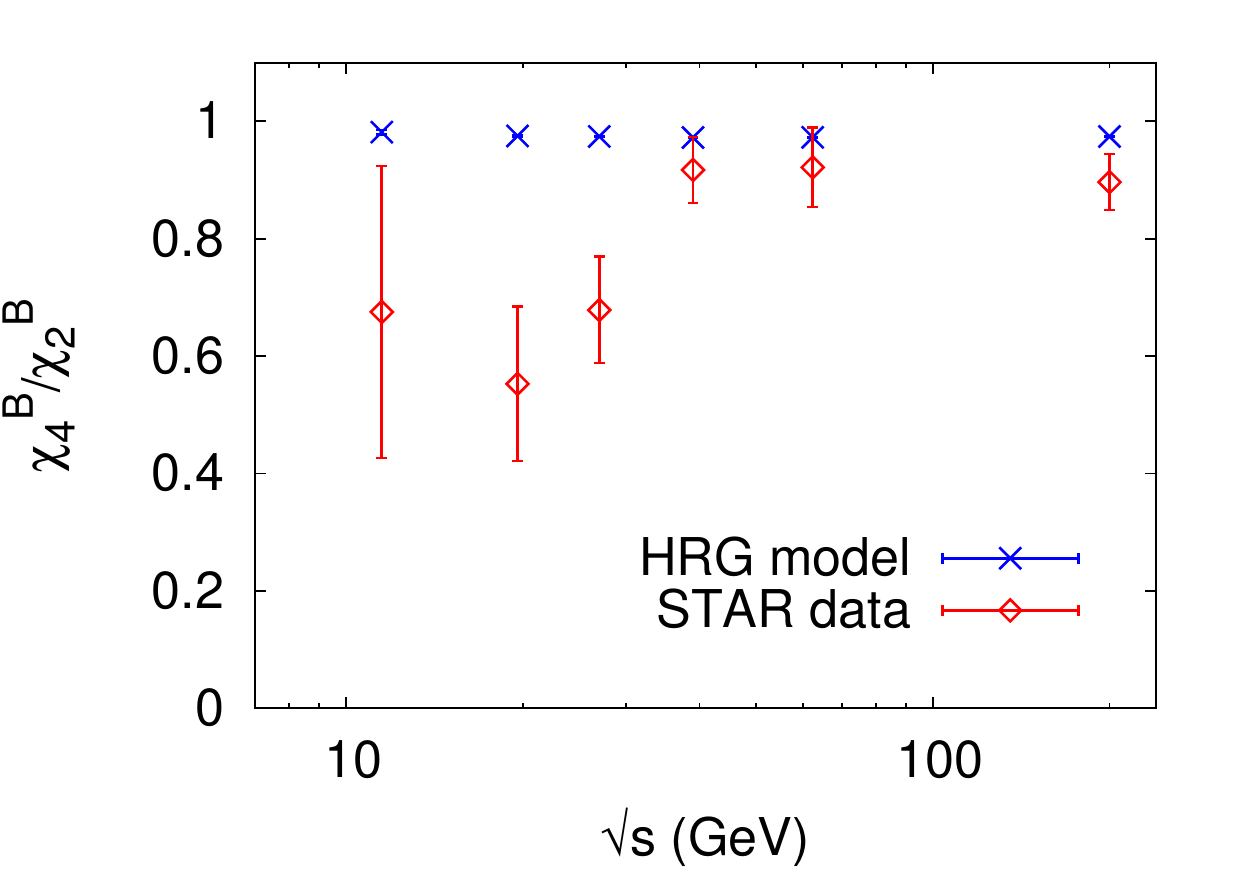}
\end{minipage}
\caption[]{\label{fig4} (Color online) Comparison between HRG model results for $\chi_3^X/\chi_2^X$ and $\chi_4^X/\chi_2^X$, with $X=Q$ (left) and $X=B$ (right) as functions of $\sqrt{s}$ (blue crosses), and experimental data for the most central collisions ($0-5\%$) from the STAR collaboration~\cite{Adamczyk:2013dal,charge} (red diamonds). The HRG model results are calculated for our new freeze-out conditions, listed in Table~\ref{tab1}. In all panels, acceptance cuts in the kinematics have been introduced, following the experimental analysis.}
\end{figure}

Using these freeze-out conditions, we now proceed to calculate the higher-order susceptibility ratios $\chi_3/\chi_2$ and $\chi_4/\chi_2$ for net protons and net-electric charge.
The results are shown in the different panels of Fig.~\ref{fig4} in comparison with the experimental data. It is evident that, with the obtained freeze-out conditions, one can reproduce all experimental results for the net-electric charge fluctuations (left panels). As already mentioned, the agreement between our results and the experimental data for the net-proton $S\sigma$ becomes less accurate with decreasing collision energy (upper right panel). For $\kappa\sigma^2$, our HRG model cannot reproduce the anomalous depletion at the lower collision energies (lower right panel), but is in good agreement with the data for the very low and very high $\sqrt{s}$ (notice that this depletion disappears in more peripheral collisions and can be described in central collisions by uncorrelated, i.e. independent, particle production when the experimentally determined proton and anti-proton distributions from STAR are used~\cite{Adamczyk:2013dal}).

\begin{figure}[h!]
\centering
\begin{minipage}{0.5\textwidth}
\includegraphics[width=1\textwidth]{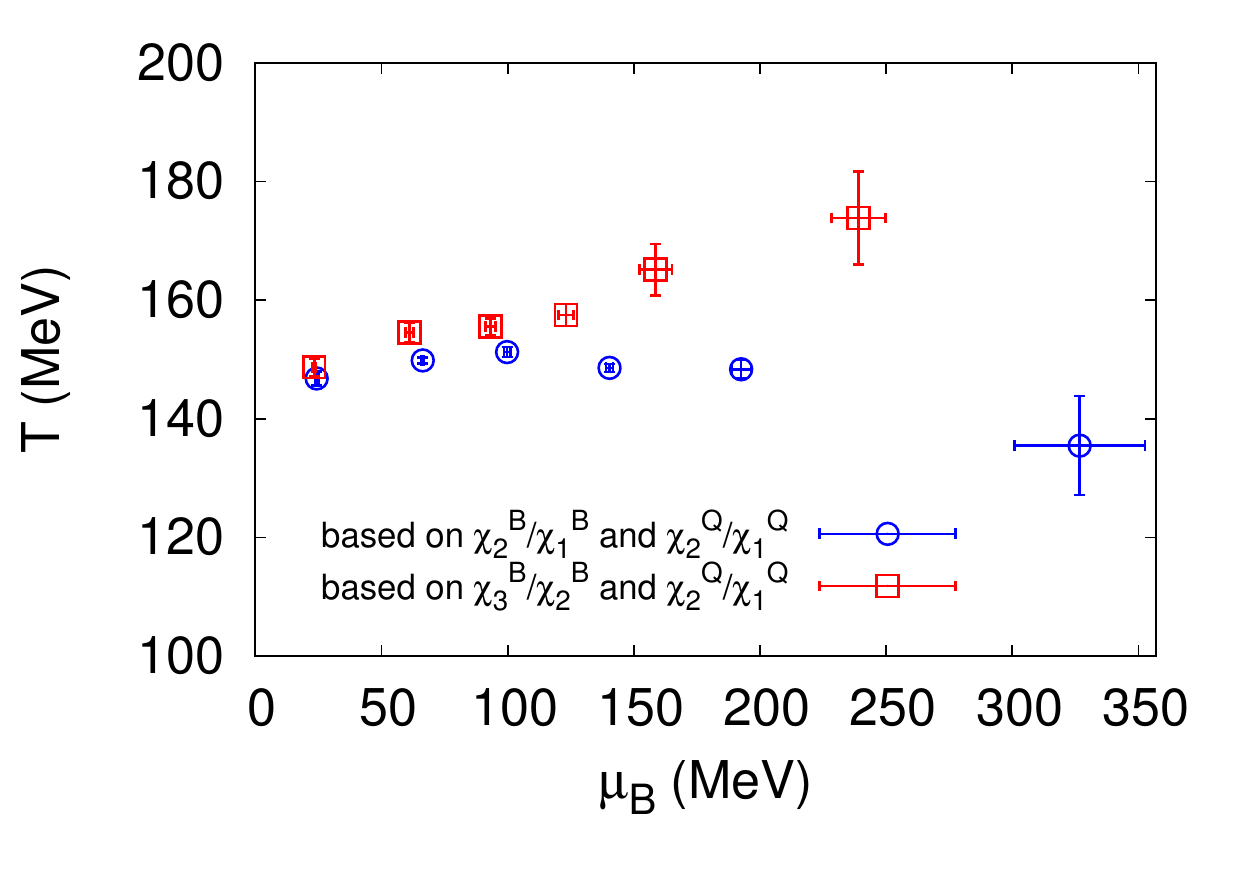}
\end{minipage}
\caption{\label{fig6} (Color online) Freeze-out parameters in the ($T-\mu_B$) plane: comparison between the values obtained by a combined analysis of $\sigma^2/M$ for net-electric charge and net protons (blue circles), and the values obtained from $\sigma^2/M$ for net-electric charge and $S\sigma$ for net protons (red squares).}
\end{figure}

In order to determine by how much the freeze-out conditions need to be modified to reproduce the higher-order cumulants for the net protons, we perform, as second choice, a simultaneous analysis of $\sigma^2/M$ for the net-electric charge and $S\sigma$ for the net protons. The result improves the agreement with the measured $S\sigma$, the values for $\sigma^2/M$ of the net-proton distributions are, however, not described particularly well. For the net-electric charge fluctuation data the uncertainties are such that the alternate analysis is still within experimental error bars.

The comparison between the freeze-out parameters resulting from our two different calculations is shown in Fig.~\ref{fig6}. While for high collision energies the two parameter sets are very similar, differences arise for smaller $\sqrt{s}$. In order to be able to reproduce the higher-order cumulants, the curvature of the freeze-out curve turns up for larger baryo-chemical potential, which is opposite to lattice expectations. This points at an inconsistency between the HRG model description of the lower- and higher-order cumulants in the net-proton distributions, which might signal the onset of chiral critical fluctuations in the higher order cumulants at large $\mu_B$~\cite{Redlich:2012xf,Skokov:2011rq}. In order to test whether, in the net-proton analysis, already the second order cumulant could be affected, we have determined the $\sigma^2/(\langle N_p\rangle + \langle N_{\bar{p}}\rangle)$ ratio and found it to be consistent with unity for both, the HRG model and the data, for all measured collision energies. In addition, since the gross features of the particle distributions are given by their lower-order cumulants, while higher-order cumulants are more sensitive to finer details, such as excluded volume effects or volume fluctuations, as well as to interactions in the late hadronic stage, obtaining chemical freeze-out parameters from the analysis of $\sigma^2/M$ for net-electric charge and net-proton number is more reliable than using $S\sigma$ for the net-proton number. The corresponding values for the freeze-out temperature and baryo-chemical potential for the different collision energies are given in Table~\ref{tab1}.

\begin{table}[h!]
\vspace{5mm}
\centering
 \begin{tabular}[t]{|c|c|c|}
 \hline
 $\sqrt{s}$ [GeV] & $\mu_{B,ch}$ [MeV] & $T_{ch}$ [MeV] \\
 \hline
 11.5 & 326.7$\pm$25.9 & 135.5$\pm$8.3\\
 19.6 & 192.5$\pm$3.9 & 148.4$\pm$1.6\\
 27 & 140.4$\pm$1.4 & 148.5$\pm$0.7\\
 39 & 99.9$\pm$1.4 & 151.2$\pm$0.8 \\
 62.4 & 66.4$\pm$0.6 & 149.9$\pm$0.5 \\
 200 & 24.3$\pm$0.6 & 146.8$\pm$1.2\\
 \hline
 \end{tabular}
 \caption[]{\label{tab1} In this table we list the values of $\mu_{B,ch}$ and $T_{ch}$ at chemical freeze-out, corresponding to the relative collision energies. These values are based on our combined analysis of the data in Fig.~\ref{fig1}.}
\end{table}

\section*{Conclusions}

In conclusion, our study shows that we can simultaneously describe the net-electric charge fluctuations and the lower-order cumulants of the net-proton multiplicity distributions measured at RHIC for collision energies spanning over more than an order of magnitude ($\sqrt{s}=(11.5 - 200)$~GeV). We calculated these fluctuation observables within the HRG model including the experimental acceptance cuts and the effects of resonance decays and regeneration.

From a combined analysis of $\sigma^2/M$ for net-electric charge and net-proton number, we obtain the freeze-out conditions summarized in Table~\ref{tab1}. Given the reported uncertainties in the measured fluctuation observables, the resulting freeze-out temperatures are constrained to better than 5 MeV for $\sqrt{s}>11.5$~GeV. With these freeze-out values, the higher-order susceptibility ratios for net-electric charge and net-proton number are reasonably well reproduced. If one takes the experimentally given particle samples as approximate representatives for the quantum numbers of electric and baryon charge, similar studies in lattice QCD yield a remarkable agreement for the collision energy dependence of $T_{ch}$ and $\mu_{B,ch}$~\cite{xxx}.

\begin{figure}[h!]
\centering
\begin{minipage}{0.5\textwidth}
\includegraphics[width=0.9\textwidth]{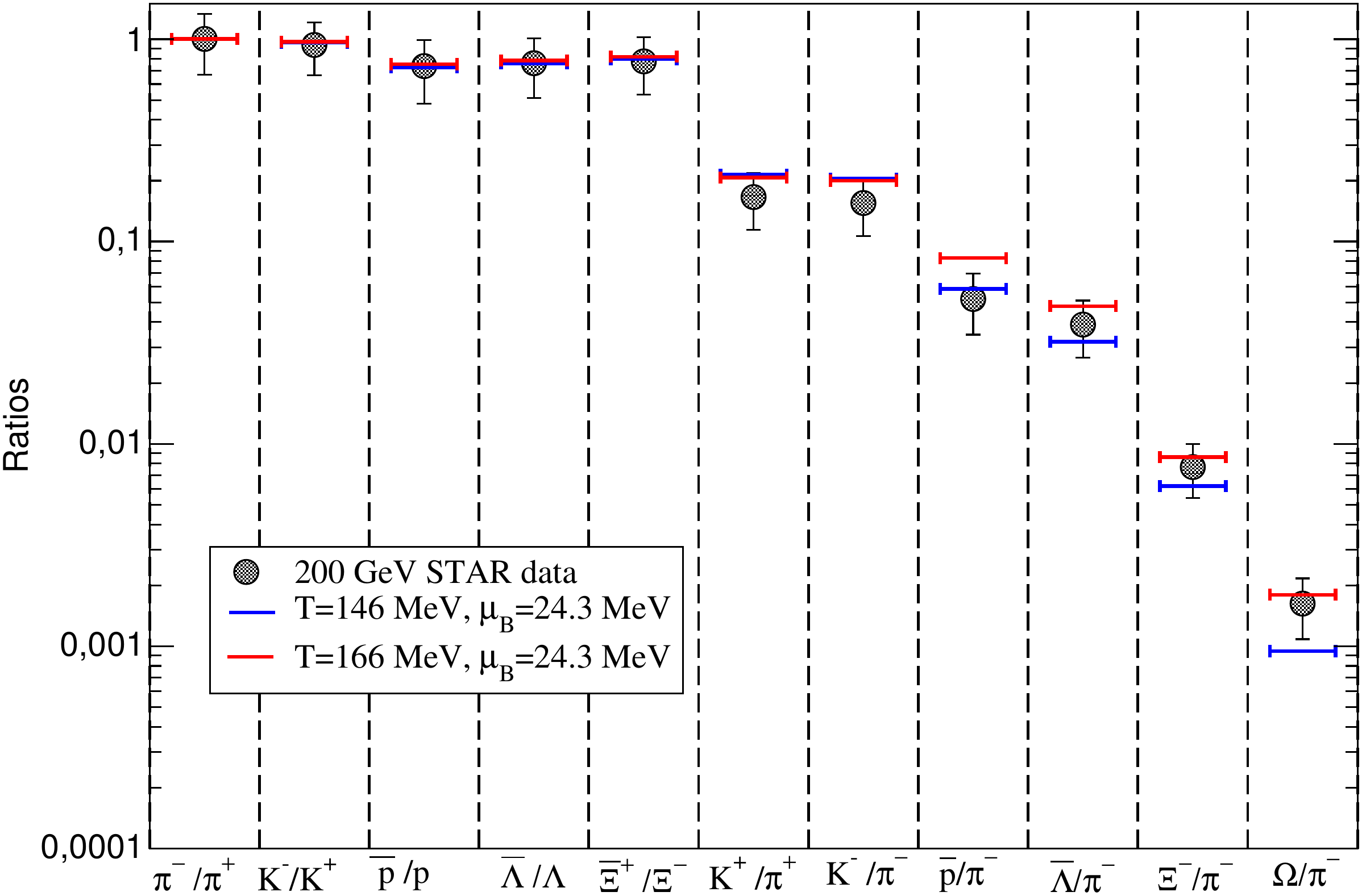}
\end{minipage}
\caption{\label{fig7} (Color online) Comparison between STAR particle ratio data for central events at $\sqrt{s}$=200 GeV~\cite{Andronic:2012dm} and HRG model results for the specified chemical freeze-out parameters.}
\end{figure}

We note that a useful cross-check of our extracted chemical freeze-out parameters can be provided through the independent determination of the same parameters via a common fit of standard SHMs to experimental particle yields or ratios~\cite{Cleymans:2005xv,Becattini:2005xt,Andronic:2005yp}. Preliminary results of measured particle ratios from the RHIC beam energy scan~\cite{Das:2014kja} for all $\mu_B$ values analyzed here, yield, based on those standard SHM fits, freeze-out temperatures ranging from ($140 - 160$)~MeV for $\sqrt{s}=(7.7 - 200)$~GeV collisions when using a common fit for all particles (including strange particles).
At first glance, our parameters are below those extracted from the particle ratio fits as is also evident from Fig.~\ref{fig3}. In order to quantify this difference we show in Fig.~\ref{fig7} a comparison of particle ratios obtained with our parameters and some standard SHM parameters to the properly feed-down corrected particle ratios measured by STAR at the highest RHIC collision energy~\cite{Andronic:2012dm}.

\vspace{0.2cm}

The overall agreement with the data is roughly equivalent for both parameter sets. Our calculation significantly improves on the (anti-)proton yield, but falls short for the strange baryons. We note that the cumulant ratios of net charge and net protons are pion and proton dominated, which could be the reason that in our analysis the particle and fluctuation ratios yield consistently lower temperatures for the light quark sector. Recent LHC data~\cite{Abelev:2012wca,Preghenella:2011np} seem to suggest a separation of chemical freeze-out parameters according to particle flavor, which is also supported by the latest lattice QCD simulations~\cite{Bellwied:2013cta} and sequential SHMs~\cite{Bugaev:2013fsa}. A fit to the strange baryon over pion ratios alone yields chemical freeze-out temperatures that are consistent with the standard SHM fits ~\cite{Preghenella:2011np}. In order to determine the freeze-out parameters from fluctuation ratios in the strange sector we anticipate that at least efficiency corrected strange meson cumulants from the experiments will be available soon.

\section*{Acknowledgements}

We gratefully acknowledge useful discussions with Bill Llope. This work is supported by the Italian Ministry of Education, Universities and Research under the Firb Research Grant RBFR0814TT, the Hessian LOEWE initiative Helmholtz International Center for FAIR, and the US Department of Energy grants DE-FG02-03ER41260, DE-FG02-05ER41367 and DE-FG02-07ER41521.

\end{document}